\documentclass[aps, prl, twocolumn, superscriptaddress, longbibliography]{revtex4-2}
\usepackage{hyperref}
\usepackage{amsfonts}
\usepackage{amsmath}
\usepackage{amssymb}
\usepackage{graphicx}
\usepackage{braket}
\usepackage{enumitem}
\usepackage{xcolor}
\usepackage{xifthen}

\newcommand{\beq}{\begin{equation}}
\newcommand{\eeq}{\end{equation}}

\newcommand{\beqa}{\begin{eqnarray}}
\newcommand{\eeqa}{\end{eqnarray}}

\newcommand{\ketbra}[2]{\ensuremath{\ket{#1}\!\bra{#1}}}
\newcommand{\Sstate}[0]{\ensuremath{4\mathrm{S}_{1/2}}}
\newcommand{\Pstate}[0]{\ensuremath{4\mathrm{P}_{3/2}}}
\newcommand{\Dstate}[0]{\ensuremath{3\mathrm{D}_{3/2}}}

\begin{document}

\title{Single-ion optical autocorrelator}

\author{M. I. Hussain}
\email{m.hussain@lmu.de}
\affiliation{Institute for Quantum Optics and Quantum Information (IQOQI), Austrian Academy of Sciences, Innsbruck 6020, Austria}
\affiliation{Institute for Experimental Physics, University of Innsbruck 6020, Austria}
\affiliation{Ludwig-Maximilians-Universität München, Am Coulombwall 1, 85748 Garching, Germany}
\author{M. Guevara-Bertsch}
\affiliation{Institute for Quantum Optics and Quantum Information (IQOQI), Austrian Academy of Sciences, Innsbruck 6020, Austria}
\affiliation{Institute for Experimental Physics, University of Innsbruck 6020, Austria}
\author{E. Torrontegui}
\affiliation{Departamento de F\'{\i}sica, Universidad Carlos III de Madrid, Avda. de la Universidad 30, 28911 Legan\'es, Spain}
\affiliation{Instituto de F\'{\i}sica Fundamental IFF-CSIC, Calle Serrano 113b, 28006 Madrid, Spain}
\author{J. J. Garc{\'\i}a-Ripoll}
\affiliation{Instituto de F\'{\i}sica Fundamental IFF-CSIC, Calle Serrano 113b, 28006 Madrid, Spain}
\author{R. Blatt}
\author{C. F. Roos}
\email{christian.roos@uibk.ac.at}
\affiliation{Institute for Quantum Optics and Quantum Information (IQOQI), Austrian Academy of Sciences, Innsbruck 6020, Austria}
\affiliation{Institute for Experimental Physics, University of Innsbruck 6020, Austria}

\begin{abstract}
Well-isolated quantum systems are exquisite sensors of electromagnetic fields. In this work, we use a single trapped ion for characterizing chirped ultraviolet (UV) picosecond laser pulses. The frequency-swept pulses resonantly drive a strong dipole transition via rapid adiabatic passage, resulting in near-deterministic population exchange caused by absorption or stimulated emission of photons. When subjecting an ion to counterpropagating pulse pairs, we observe the loss and revival of atomic coherence as a function of the pulse pair spatial overlap---enabling quantification of the temporal pulse broadening caused by a frequency chirp in shaped UV pulses with a very low peak power. We find good agreement between measured and applied chirp. The ultrafast population exchange imparts an impulsive force where the estimated change in the mean phonon numbers of ~$\sim0.5$ is measured for two pairs of pulses. The resonant ultrafast kicks could be applied to matter-wave interferometry experiments and present a step towards ultrafast entanglement operations in trapped ions.

\end{abstract}  	
\maketitle

Recent progress in optical science has tremendously leveraged our understanding of the nature of light, matter, and in general light-matter interaction. Particularly, exciting matter with ultrafast lasers provides insights into physical processes at extremely short timescales ranging from pico to attoseconds \cite{fermann2002ultrafast,keller2003recent,calegari2014ultrafast,chini2014generatio}. 
Phase engineering across the entire spectrum of the laser pulse, so-called frequency chirping \cite{weiner2011ultrafast} opens new avenues for coherent control \cite{goswami2003optical}, which includes applications in molecular spectroscopy \cite{leahy2010ultrafast}, optical communications \cite{agrawal1986effect}, synthesis of photochemical processes \cite{assion1998control}, and chirped pulse amplification \cite{strickland1985compression,maine1988generation}. Interferometry based on chirped pulses benefits atomic hydrogen spectroscopy for quantum electrodynamics \cite{grinin2020two}, steering molecular dynamics, \cite{bardeen1995selective,pe2007precise}
wavepacket interferometry \cite{bouchene1999wavepacket}, to name a few.

Moreover, pulse-shaping allows for efficient population inversion \cite{loy1974observation,loy1978two,oberst2007efficient} and pumping-dumping control through rapid adiabatic passages (RAPs) \cite{malinovsky2001general} that can be applied to atomic and molecular acceleration/deceleration \cite{miao2007strong,jayich2014continuous,long2019suppressed}. The philosophy behind most of such investigations connects with the underlying Landau–Zener (LZ) physics \cite{landau1932theorie,zener1932non}, which, with the advent of modern quantum technology, has been studied in a wide variety of different qubit systems \cite{oliver2005mach,Wunderlich:2007,fuchs2011quantum,forster2014characterization} such as superconducting qubits, trapped ions, NV centers or quantum dots.

Laser-manipulated trapped ions provide a well-controlled quantum system with control over internal and motional degrees of freedom \cite{leibfried2003quantum} that has led to many applications in quantum computing, simulation and sensing \cite{Ludlow:2015,Bruzewicz:2019,Monroe:2021,blums2018single,Wolf:2021}. However, simultaneously achieving excellent coherent control and high-speed manipulation of the atomic motion has remained elusive so far. The motivation to realize this goal has driven the development of ultrashort pulse techniques for trapped-ion experiments \cite{campbell2010ultrafast,mizrahi2013ultrafast,hussain2016ultrafast,johnson2017ultrafast,Wong-Campos:2017,hussain2021ultraviolet,shinjo2021three,shimizu2021ultrafast, guo2022picosecond,Putnam:2023} that can be beneficial to many of the above mentioned applications by reducing the time needed for quantum state manipulation. In these experiments, single picosecond pulses or pulse trains of picosecond pulses were employed for creating $\pi$-pulses, spin-motion entanglement and demonstrating two-ion entanglement. As compared to resonant $\pi$-pulses, chirped pulses enable the realization of RAPs, which offer increased robustness to laser intensity fluctuation. Chirped pulses have been applied to trapped ions \cite{Wunderlich:2007,watanabe2011sideband,Lechner:2016} on dipole-forbidden atomic transitions, but only recently have chirped picosecond pulses been employed for realizing RAPs on atomic dipole transitions \cite{Song:2020,hussain2021ultraviolet}.

In this letter, we show that a pulse pair obtained from a single chirped optical pulse can coherently drive the population between the ground and the excited states of a trapped ion via RAPs on a picosecond time scale.
We steer pulses on the ion in a counterpropagating direction for successive population inversions in a strong dipole transition.
In this way, we achieve a stimulated emission via resonant UV optical pulses---and accelerate the ion with momentum transfer of~2$\hbar$k in as little as $\sim$6 ps. This paves the way for ultrafast quantum control and overcomes a major hurdle in the implementation of fast entangling gate operations in trapped-ion quantum processors.\cite{garcia2003speed}.
We exploit these novel experimental tools for sensing ultrafast laser pulses. Specifically, we present a proof-of-concept study demonstrating a single-ion autocorrelator that can serve as a versatile tool for \textit{in-situ} diagnostics of narrowband, low peak power, and chirped ultrafast UV pulses derived from a very low-duty cycle optical signal---i.e., picking a single pulse out of a pulse train with 5 GHz repetition frequency and the optical signal period of 1 ms \cite{hussain2021ultraviolet}.
\begin{figure}
\includegraphics[width=8cm,height=4.2cm]{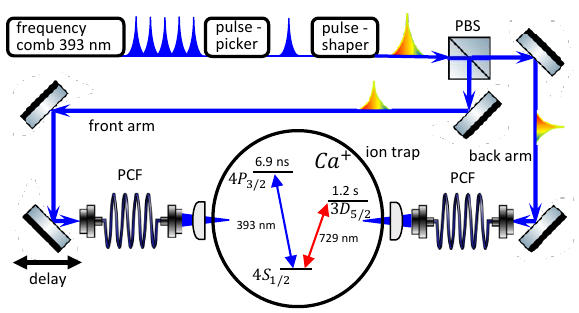}
\caption{Schematic of the experiment. Picosecond 393 nm laser pulses pass through the pulse picker and subsequently travel through the pulse shaper for a frequency chirp. Next, the pulses are split and focused onto a single Ca$^+$ ion trapped in a Paul trap. PCF: photonic crystal fiber, PBS: polarization beam splitter.}
\label{setup}
\end{figure} 
In this case, common pulse characterization \cite{trebino2000frequency} methods struggle due to small peak intensities and inefficient nonlinear mixing as a result of a difficult operational wavelength. More sophisticated techniques for ultraweak pulses e.g., spectral interferometry \cite{fittinghoff1996measurement} require the spectrum of the unknown pulse within the reference pulse's spectral region and reference-pulse monitoring with frequency-resolved optical gating, thus making it inconvenient as well. These difficulties are overcome by characterizing the pulses with the same system on which they will be used \cite{cadarso2014phase,ding2021measuring}, which we achieve by subjecting an ion to a pair of partially overlapping counterpropagating pulses.

Our experiments are conducted with a linear Paul trap housed in an ultra-high vacuum system, as detailed in \cite{guggemos2015sympathetic}. Permanent magnets create a magnetic bias field aligned with the trap's symmetry axis.
Single Ca$^+$ ions are loaded into the trap by photoionization of laser-ablated atoms.  Doppler cooling and state detection are achieved on the $4\mbox{S}_{1/2}$ and $4\mbox{P}_{1/2}$ dipole transition with a diode laser frequency-doubled to 397 nm. 
Population from the metastable $3\mbox{D}_{3/2}$ and $3\mbox{D}_{5/2}$ states is repumped with 866 nm and 854 nm light respectively \cite{guggemos2015sympathetic}.
The coherent manipulation of the quadrupole clock transition $4\mbox{S}_{1/2}\leftrightarrow 3\mbox{D}_{5/2}$ (qubit states) is achieved by an ultrastable 729 nm Ti-Sapphire laser with a linewidth of less than 10 Hz.

For picosecond coherent excitation of the $4\mbox{S}_{1/2}\leftrightarrow 4\mbox{P}_{3/2}$ transition, a mode-locked frequency comb generates telecom-band pulses followed by repetition rate upconversion (250 MHz to 5 GHz) and frequency quadrupling from 1572 nm to a wavelength of $\lambda=393$~nm. 
The pulse trains are patterned by fast electro-optical switches, which enable us to extract individually switched pulses (see Fig.~\ref{setup}); for further details, see ref.~\cite{hussain2021ultraviolet}.
Subsequently, 393 nm pulses are allowed to pass through a grating pair, where we add a group delay dispersion (GDD) for a frequency chirp. In this way, the initially $\sim$1.5 ps long UV pulses are stretched, depending on the amount of GDD. A polarization beam splitter splits chirped pulses, which are then coupled into two 2~m long photonic crystal fibers. These beam paths constitute the arms of an interferometer that is closed by the action of a pair of counter-propagating pulses on the trapped ion.
The path length in one of the arms is adjustable via a delay line to adjust the arrival time difference between counter-propagating pulse pairs. Linearly polarized pulses are directed along the trapping axis and are focused on the ion through the endcap electrodes to a beam waist of 6 $\mu$m.

\begin{figure}

\includegraphics[width=8.5cm,height=5.5cm]{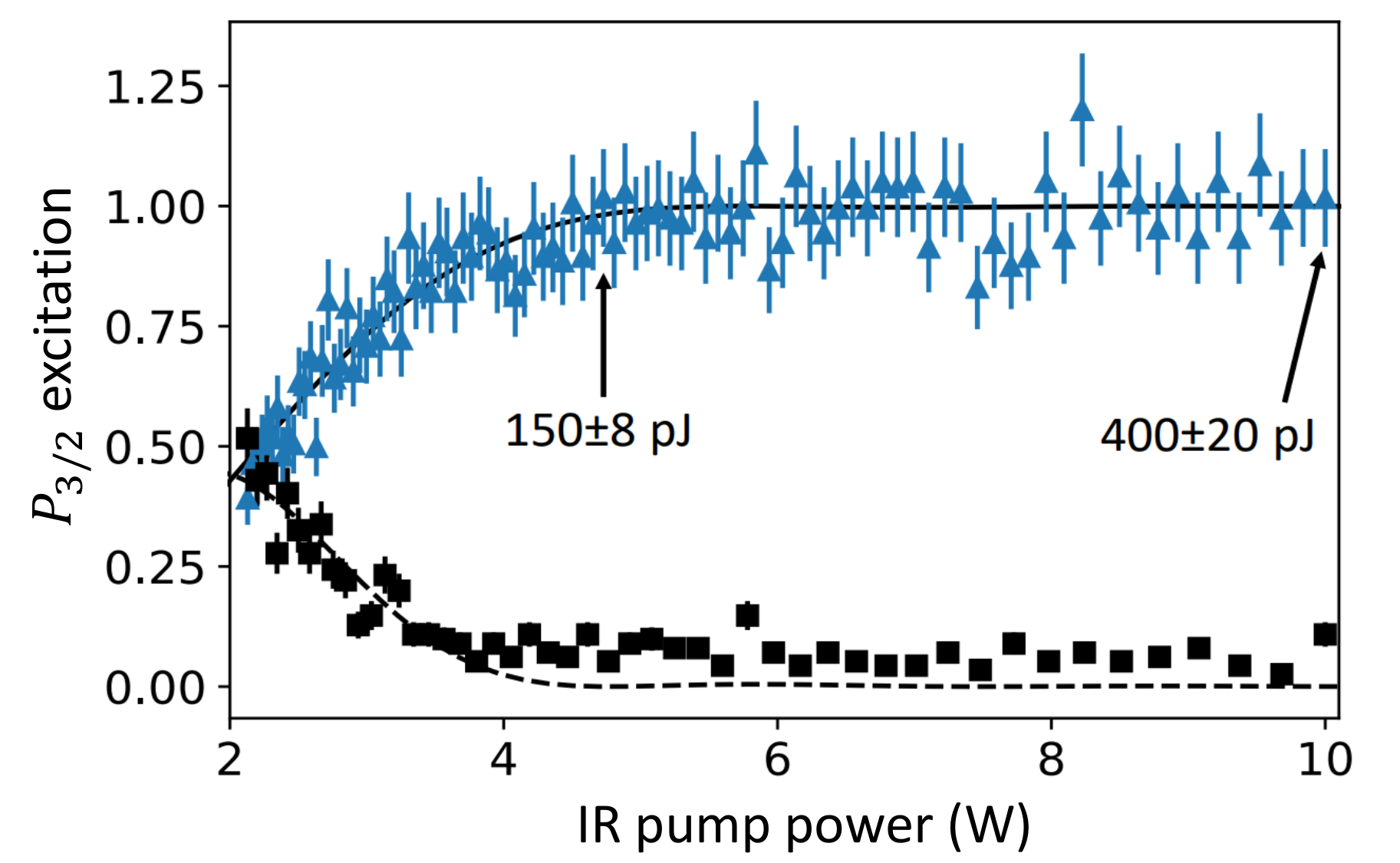}
\caption{(a) Demonstration of rapid adiabatic passages. A trapped ion is exposed to a single (triangle, blue) and a pair of counterpropagating (square, black) pulses whose pulse energy is scanned by increasing the pump power at 1572 nm. Black arrows indicate pump powers for which the corresponding indicated single-pulse energies of the 393 nm laser pulses are high enough to maximize the population transfer probability. The high-power pulse amplifier's tunability limits the scanning range in the fundamental pump power. For this reason, the scanning power and excitation counts start from nonzero values. Black-solid and black-dashed lines are numerical simulations of single and double RAPs. The unphysical excitation probabilities $>$ 1 are statistical measurement fluctuations that result from inferring the upper-state population from the fraction of the population having decayed to the $4\mbox{D}_{5/2}$ state using a finite number of experimental repetitions (cf. ref.~\cite{heinrich2019ultrafast}).}
\label{fig:rap}
\end{figure}

In a first experiment, we measure the pulse energy required for realizing a RAP on the $4\mbox{S}_{1/2} \leftrightarrow 4\mbox{P}_{3/2}$ transition by exciting a Doppler-cooled ion prepared in the state $|S\rangle\equiv|4\mbox{S}_{1/2}, m_J = 1/2\rangle$ to the state $|P\rangle\equiv|4\mbox{P}_{3/2}, m_J = 3/2\rangle$, using a chirped pulse (obtained by blocking one of the beam paths) or a pair of chirped pulses, respectively. We infer the excitation probability at the end of the interaction by measuring the population having decayed into the $4\mbox{D}_{5/2}$ \cite{heinrich2019ultrafast} as a function of the pulse energy. Figure~\ref{fig:rap} shows that for pulse energies above 200~pJ, a single chirped pulse inverts the population whereas a chirped pulse pair (CPP) separated by $\tau=7$~ps coherently excites and de-excites the ion with high probability.

We can describe the single-pulse RAP dynamics with a time-dependent effective Hamiltonian
\beq
\label{heff}
H_\text{eff}(\tau) =-\frac{\delta^2\tau}{2}\sigma^z + \Omega_D w(\tau) \sigma^x
\eeq
with the Pauli spin matrices $\sigma^z = |P\rangle\!\langle{P}|-|{S}\rangle\!\langle{S}|$ and $\sigma^x = |S\rangle\!\langle{P}|-|{S}\rangle\!\langle{P}|$.
The internal atomic states interact with a chirped Gaussian field $w(\tau)=\exp(-\tau^2/(2\sigma_D^2))$, with amplitude $\Omega_D$, relative time $\tau=t-x/c$, stretched width $\sigma_D$ and chirp factor $\delta$.

As discussed in the appendix \cite{SM}, the dynamics under two RAP pulses can be computed from the unitary evolution under a single pulse, with some phase delay $\phi$. When $p_1$ is the excitation probability under one pulse [cf. Triangles and solid line in Fig.~\ref{fig:rap}], the probability of residual excitation after two consecutive pulses is $p_2 = 4 \braket{\cos(\phi/2)^2} p_1 (1-p_1)$. In the absence of phase fluctuations, when $p_1=0.5$, two consecutive pulses coherently add up to a $p_2=1$ excitation. In the experiment, however, the incoherent averaging over all possible phases $\braket{\cos(\phi/2)^2}=1/2$ produces a crossing between both experiments at $p_1=p_2=0.5$, as shown in Fig.~\ref{fig:rap}.

\begin{figure}
\includegraphics[width=8.2cm,height=6.5cm]{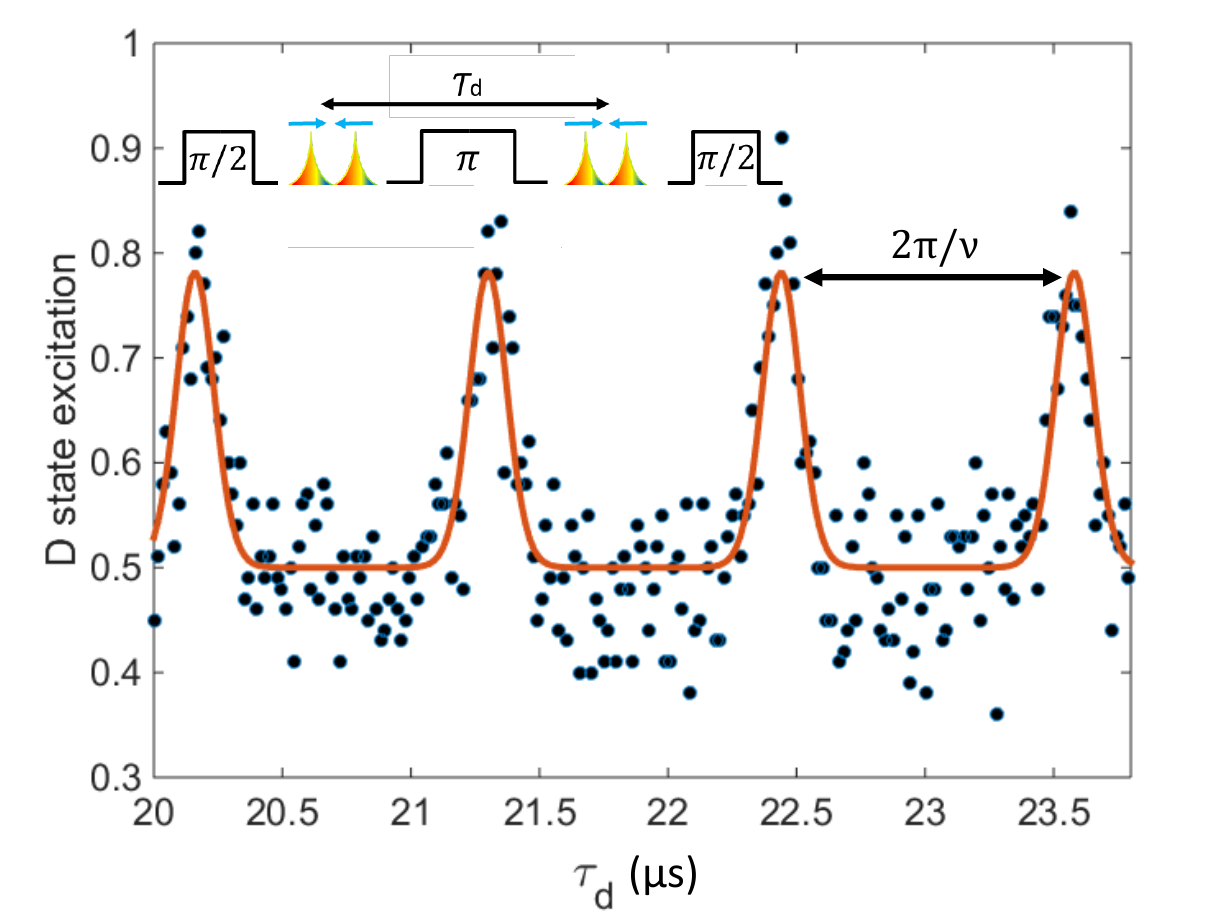}
\caption{Investigation of CPPs with counter-propagating pulses. Two CPP pulse pairs with variable delay $\tau_d$ are embedded into a spin-echo sequence, as shown in the inset. Here and anywhere else in the paper, blue arrows indicate the propagation direction of the pulses with respect to the ion. The probability to excite the ion to the D-state oscillates as a function of the inter-CPP delay $\tau_d$ with a period equal to the inverse of the ion's secular frequency $\nu = 2\pi\times 890$~kHz. Maximum excitation is achieved whenever both CPPs coherently displace the $|S\rangle$ and $|D\rangle$ state populations in the same direction in phase space such that they can interfere at the end of the spin echo sequence. Circles represent measured data, and the solid line is fit with a spin echo contrast of $C_0=0.56(2)$ and a thermal state with $\bar{n}=21 (2)$ motional quanta.}
\label{delay}
\end{figure}

While two consecutive RAPs leave the electronic state of an ion prepared in $\mbox{S}_{1/2}$ invariant, they can nevertheless coherently transform the ion's quantum state. Firstly, double RAPs realized by a counterpropagating CPP transform the electronic state via $|S\rangle\rightarrow e^{i\theta}|S\rangle$ with a rotation angle $\theta$ determined by the path length difference of the two interferometer arms. This phase factor matters if the ion was initially in a coherent superposition of $|S\rangle$ and another (meta-)stable electronic state. Secondly, each RAP changes the ion's motional state by coherently displacing it in phase space by an amount $\eta$, where $\eta$ is the Lamb-Dicke factor of the $|S\rangle\leftrightarrow|P\rangle$ transition. Therefore, for an ion initially prepared in a superposition of electronic states, a RAP creates quantum correlations between internal and motional states. 

To investigate these effects, we carry out Ramsey and spin echo experiments on the transition connecting the $|S\rangle$ state to the long-lived $|D\rangle\equiv|\mbox{D}_{5/2},m=3/2\rangle$ state and inserted CPP's into the free evolution times. To evaluate the fringe contrast, we measure the D-state population as a function of the phase of the last $\pi/2$ pulse. For a Ramsey experiment, in which both pulses of the CPP travel along the same path, the contrast stays above 90\%, slightly lower than the 98(2)\% contrast observed without the CPP. However, the contrast completely vanishes for a CCP made of counterpropagating pulses, which is expected as the relative path length difference drifts by more than $\lambda/2$ over the time scale of a second.

A high fringe contrast can be recovered by sandwiching counterpropagating CPP's into each of the two free evolution periods of a spin echo sequence. In this case, both electronic states pick up the same phase factor, provided that the optical path length is stable over the duration of the spin echo. Additionally, both CPP's displace the motional state in phase space by $2\eta$ in the same direction. However, the harmonic trapping potential rotates any displaced state by an angle $\theta=\nu\tau_d$ where $\nu$ is the axial oscillation frequency, $\tau_d$ the time after the displacement operation $\hat{D}$. Therefore, after the second CPP, the $|S\rangle$ and $|D\rangle$ states have been displaced by $\hat{D}(2\eta e^{i\theta})$ and $\hat{D}(2\eta)$, respectively and the displacement becomes state-independent when $\theta = 2\pi m$ for integer values of $m$. For other delays, the state-dependent displacements lead to a reduction of contrast \cite{mizrahi2013ultrafast,guo2022picosecond,Putnam:2023}, which for a thermal state with mean phonon number $\bar{n}$ varies as $C(\tau_d)=\frac{1}{2}(1+C_0e^{-|\alpha(\tau_d)|^2(\bar{n}+1/2)}\cos(4\eta^2\sin\nu\tau_d))$ with $\alpha(\tau_d)=8\eta^2(1-\cos\nu\tau_d)$ where $C_0<1$ accounts for experimental imperfections of the spin echo and the CPP's.

The data shown in Fig.~\ref{delay} was taken with the phase of the last $\pi/2$ pulse set to maximize the excitation in the absence of any CPP's. The measured signal displays the predicted time-periodic variation, with the peaks corresponding to pulse pairs separated in time by 18 to 21 trap oscillation periods. Fitting the data with the contrast function $C(\tau_d)$, we obtain $\bar{n}=21(2)$ which is consistent with the temperature of a Doppler-cooled ion measured by sideband spectroscopy.

\begin{figure}
\includegraphics[width=8.2cm,height=5.2cm]{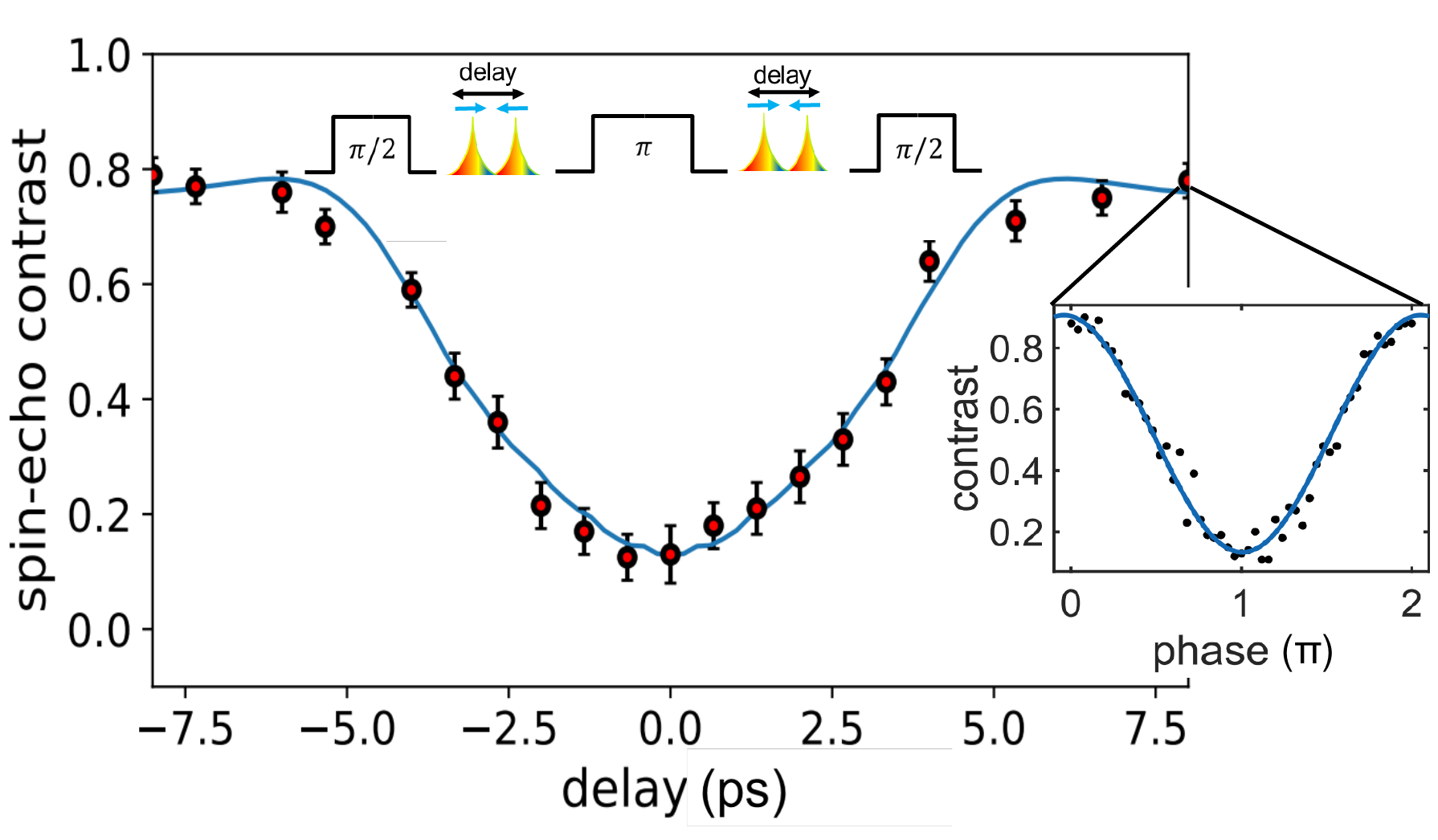}
\caption{
Single-ion autocorrelation measurements obtained by embedding two CPP's into the spin echo sequence shown at the top of the figure. We measure the spin echo contrast as a function of the arrival time delay $T_D$ between two pulses impinging on the ion from opposite sides, with the inter-CPP separation $\tau_d$ set to a value that enables observation of the maximum contrast. As shown in the inset, a high fringe contrast is observed whenever $T_D$ is substantially bigger than the pulse width of the pulses. When the pulses start to overlap, the spin echo contrast collapses. The solid blue line represents a fit to the data, based on a numerical simulation that is used to estimate the pulse length of the chirped pulses. All data points are taken at comparatively low pulse energies ($\le150$ pJ) that, however, are still high enough to realize a RAP in the case of non-overlapping pulses.
}
\label{AC}
\end{figure}

The embedding of double CPPs into a spin-echo sequence that is insensitive to the ion motion creates the basis for chirped pulse characterization measurements. Unlike a standard pulse autocorrelator based on nonlinear frequency conversion, we measure the short-pulse duration by an ion subjected to two copies of the pulse whose temporal overlap we vary. As soon as the pulses start to overlap, they stop acting as a RAP which leads to a loss of spin echo contrast as demonstrated in Fig.~\ref{AC}. For these measurements, we employ the same pulse sequence as for the data of Fig.~ 
\ref{delay}, but fix the temporal separation between the two pulse pairs to a value that maximizes the spin echo contrast. Moreover, we prepare the ion close to the ground state of its axial motion by sideband cooling in order to make the sequence more robust against potential trap frequency fluctuations. 

We extract the change in contrast as a function of the arrival time separation $T_D$ between the two pulses comprising a CPP at the location of the ion. Each data point is obtained by repeating the experimental cycle 200 times. The contrast value varies between a maximum of 0.8(0.035) when the pulses of the CPP are non-overlapping and a minimum of 0.13(0.045) when the two pulses arrive simultaneously. 
Note that the loss of contrast is insensitive to the sign of the time-of-arrival difference between CPP pulses. 

To analyze the autocorrelator signal (spin-echo contrast decay and revival data), we developed an ab-initio model to describe the excitation of the dipole transition $\Sstate\leftrightarrow \Pstate$ under the two pairs of pulses, both in the CPP-overlapping and -non-overlapping regimes. We fit this model numerically against the experiment, using as fit parameters the pulse's intensity $I$, the overall loss of contrast due to motional interference $\eta$, the initial pulse width $\sigma$, and the effective GDD factor $D$. We expect the width of the loss-of-contrast profile to be related to the actual width of the chirped pulse, $\sigma_D$, which is linked to the initial pulse width $\sigma$ via $\sigma_D^2 = (\sigma^4+D^2)/\sigma^2$ (cf. \cite{SM}). Unlike conventional methods of pulse autocorrelation, the ion's excitation curve exhibits a nonlinear relation to the pulse shape that cannot be accounted for by a simple deconvolution factor. Indeed, depending on the pulse power, the width of the loss-of-contrast profile may be smaller, equal to, or larger than the chirped pulse width, as shown in the appendix Fig.\ref{fig:excitation-from-interference}. Despite this, the theoretical model provides decent estimates of the pulse width. Based on the pulse shaper design, if we fix the value of GDD to the estimate $D\sim5.8(0.05)
\mathrm{ps}^2$, the fit produces $I=0.05(0.02)$ (a.u.), $\sigma=1.5(0.2)\mathrm{ps}$ and $\eta=79(1)\%$, which leads to a FHWM $\sim 9.95(0.59)$ ps the pulses' stretched width and an averaged $\chi^2=0.73$ [cf. Fig.~\ref{AC}]. When we leave the value of the GDD as a fitting parameter, the fit produces $D=6.8(1.1)~\mathrm{ps}^2$, $\sigma\sim 1.54(0.21)$ ps, FWHM $\sim 11(2)$ ps, a slightly improved fit quality $\chi^2=0.67$, but an indistinguishable curve from Fig.~\ref{AC}. Note that before pulse shaping, the pulse widths were close to the Fourier limit, previously measured based on the cross-correlation frequency-resolved optical gating trace \cite{hussain2021ultraviolet}. The fits are nevertheless compatible with the calculated estimate (FWHM $\sim 10.8(0.1)$ ps) relying on the properties of the pulse shaper.


The theoretical model currently does not include other effects, such as self-phase modulation (the experiment is designed to prevent self-phase modulation by a proper choice of fibers). Moreover, we verify the absence of self-phase modulation by repeating the experiment at different pulse energies above the threshold needed for realizing RAP's with near-unit efficiency and verifying that the excitation efficiency does not change significantly.

In order to detect and quantify CPP-induced change of the ion's motional state, the ion is prepared close to the motional ground state via sideband cooling ($\langle n\rangle_0=0.08(1)$) before subjecting it to a variable number of CPPs which are separated in time by an integer multiple of the ion's oscillation period such that the individual momentum kicks coherently add up. For the analysis of the motional state, red and blue sideband excitation probabilities $p_{red}(p_{blue})$ were measured for a fixed sideband pulse length in order to determine the mean phonon number via $\langle n\rangle = p_{red}/(p_{blue}-p_{red})$. As this formula is valid only for thermal states of motion, we limited the maximum coherent displacement to an amount where the state is practically still indistinguishable from a thermal state. For the case of two CPPs, we measure a mean phonon number of $\langle n\rangle=0.58(1)$ from which we infer an increase due to the CPPs of $\Delta n=\langle n\rangle-\langle n\rangle_0=0.50(1)$. This is in good agreement of the theoretically expected value of $\Delta n_{theo}=16E_{rec}/(\hbar\nu)=0.52$ where $E_{rec}=(\hbar k)^2/(2m)$ is the photon recoil energy and $\nu$ the angular trapping frequency of the ion. If the time-separation between the two CPPs is set such that the momentum kicks of the CPPs have opposite signs, we find $\Delta n=0.04(1)$.

In this work, we explored single-ion interferometry in combination with RAPs and studied their aspects in ultrafast physics and quantum control. We have found that atomic coherence can be used to temporally characterize chirped UV pulses which are extremely difficult to measure with the standard pulse characterization techniques. The estimates in this work show an acceptable accuracy, which may be improved in future works by improving the power stability of the setup and calibrating the pulse intensity and widths in parallel experiments, working with both single-pulse excitations (e.g., Fig.~\ref{fig:rap}) and two-pulse interference. Other upgrades include the use of more sophisticated algorithms to explore the carrier-envelope phase difference~\cite{cadarso2014phase}, or using traps with more ions in magnetic field gradients to achieve single-shot pulse width estimates.

The momentum transfer as a result of a coherent population pumping-dumping enables
us to produce resonant impulsive force and strong optical force can be generated in the ultrafast time scales \cite{long2019suppressed}. Due to sufficiently high UV pulse energy, the magnitude of optical kicks can be quadrupled just by adding beam splitters in the interferometer and removing the fiber couplings. Moreover, with our short-pulse specifications a momentum transfer of $\sim 4 \times56 \hbar k$  per spontaneous emission event ($\sim$ 7 ns) seems achievable. This work investigates three crucial points in achieving high fidelity and high-speed entangling gate operations, namely, control over pulse area, pulse phase, and ultrafast motional state manipulation. In the future, our work has the potential to dramatically increase the entangling gate speed between two ions with suitable pulse sequences. To this end, modifications in the setup, e.g., complete free space pulse propagation and the interferometer's path lengths stabilization will be an imperative future step. This can potentially provide the flexibility to trigger large kicks almost at any time during the trap period and better control over the timing of the pulse kicking strategy \cite{torrontegui2020ultra}.

\begin{acknowledgments}
We acknowledge support from the Spanish Government via Projects PGC2018-094792-B-I00 and PID2021-126694NA-C22 (MCIU/AEI/FEDER,UE), CSIC Research Platform PTI-001, and by Comunidad de Madrid-EPUC3M14 and CAM/FEDER Project No. S2018/TCS-4342 (QUITEMAD-CM). E. T. acknowledges the Ram\'on y Cajal program (RYC2020-030060-I). Additionally, we acknowledge funding by the Institut f\"ur Quanteninformation GmbH.\end{acknowledgments}


\begin{center}
\textbf{ \large Appendix}
\end{center}
\renewcommand{\theequation}{A\arabic{equation}}
\renewcommand{\thefigure}{A\arabic{figure}}
\renewcommand{\citenumfont}[1]{A#1}
\section{Linear chirped pulse}
For our simulations, we assume a pulsed light source that generates Gaussian pulses $\tilde{f}_0(\omega)$ of width $\sigma$ centered around a carrier frequency $\Delta$ in the frequency domain
\beq
\tilde{\Omega}_\text{orig}(\omega) = \Omega_0 \exp\left(-\frac{1}{2}\sigma^2 (\omega-\Delta)^2\right),
\eeq
In the simulations, we assume a pulse that is resonant with the 4S$_{1/2}\leftrightarrow$ 4P$_{3/2}$ transition corresponding to transition wavelength $\lambda=393$ nm that sets a carrier frequency $\Delta=2\pi c/\lambda$.

Before interacting with the ion, this pulse experiences a frequency-domain phase delay or group delay dispersion $D$ of the form
\beq
\tilde{\Omega}(\omega) = \tilde{\Omega}_\text{orig}(\omega) \exp\left(i \frac{1}{2}D (\omega-\Delta)^2\right) = \tilde{\Omega}_\text{orig}e^{i\varphi(\omega)},
\eeq
with the group-delay dispersion (GDD) $D=\partial^2\varphi/\partial\omega^2$.
The corresponding time-dependent profile takes the form,
\beq
\Omega(t) = \frac{\Omega_0}{\sqrt{i D + \sigma^2}} \exp\left(-\frac{1}{2}\frac{t^2}{\sigma^2 - i D}-i\Delta t\right),
\eeq
that using the relation
\beq
\frac{1}{\sigma^2 - i D} = \frac{1}{\sigma^4 + D^2}(\sigma^2 + i D),
\eeq
reads,
\beq
\label{chirp}
\Omega(t) = \Omega_D \exp\left(-\frac{1}{2}\frac{t^2}{\sigma_D^2} - i\Delta t - i \frac{1}{2} \delta^2 t^2\right) = \Omega_D f(t)
\eeq
with
\beq
\sigma_D^{2} = \frac{\sigma^4 + D^2}{\sigma^2},\; \Omega_D = \frac{\Omega_0}{\sqrt{i D + \sigma^2}} \mathrm{~and~}
\delta^2 = \frac{D}{\sigma^4+D^2}.
\eeq

The chirped profile corresponds to a linear chirped Gaussian pulse with a stretched width $\sigma_D$ due to the group delay dispersion, an instantaneous frequency that increases linearly in time $\omega_{ins}(t)=\Delta+\delta^2 t$ with a carrier wave $\Delta$ and chirp factor $\delta$. In the description of the simulations and the experiments, we will typically discuss the full-width half-maximum (FWHM) size of the wavepacket's intensity. For a Gaussian profile with a ``width" parameter $\sigma_D$ as in\ \eqref{chirp}, this is defined by $|f(t_\text{FWHM}/2)/f(0)|^2=1/2$, or the equation $\exp(-\frac{1}{2}(t_\text{FWHM}/2\sigma_D)^2)=1/2$, which is solved by $t_\text{FWHM}=2\sqrt{2\log(2)}\sigma_D\simeq 2.3548 \sigma_D$. Based on this, we can also find how much the chirping stretches the FWHM as a function of the GDD and the initial width $t_\text{FWHM}(D=0)$
\begin{equation}
    t_\text{FWHM}(D) = \sqrt{\frac{t_\text{FWHM}(0)^4+ 16 D^2 \log(2)^2}{t_\text{FWHM}(0)^2}}.
\end{equation}

\section{Single-pulse effective Hamiltonian}
We consider the trapped-ion subspace corresponding to the states 4S$_{1/2}$ and 4P$_{3/2}$ that constitute the two-dimensional quantum space $\mathrm{lin}\{\ket{\Sstate},\ket{\Pstate}\}$. The Hamiltonian describing the dynamics of the ion interacting with a single coherent laser driving is
\beq
H = \frac{1}{2}\Delta\sigma^z + \omega a^\dagger a + \Omega(\tau) \sigma^+ + \Omega^*(\tau) \sigma^-,
\eeq
where $\omega\simeq 2\pi\times 1$ MHz is the harmonic trapping frequency, $\{a,a^\dagger\}$ are the Fock operators for the motion of the ion, $\sigma^z = |P\rangle\!\langle{P}|-|{S}\rangle\!\langle{S}|$, and the pulse acting on the ion
$\Omega(\tau)=\Omega_D f(\tau)$ is given by Eq.~\eqref{chirp} with an $\Omega_D$ constant field amplitude and relative time $\tau=t-x/c$. The RAP pulse width is still very short compared to the period of the ion in the electromagnetic trap. During the dynamics of the ion-pulse interaction, we can consider the ion frozen in space and treat $x$ as a c-number that offsets the time reference. When we introduce this number back, we will find $x$ appearing in a position-dependent phase shift, expressing the momentum kick the ion suffered.

The Schrödinger equation is best solved on a reference frame that rotates with pulse phase $\phi(t)=\int_0^t dt'\omega_{ins}(t')$. In this interaction picture $\ket{\Psi(t)} = \exp\left(- \frac{i}{2}\phi(\tau) \sigma^z\right)\ket{\chi(t)}$, the state $\ket{\chi}$ evolves with the effective Hamiltonian
\beq
H_\text{eff} =-\frac{\delta^2\tau}{2}\sigma^z + \Omega_D w(\tau) \sigma^x,
\eeq
with the chirped Gaussian profile $w(\tau)=\exp(-\tau^2/(2\sigma_D^2))$. In this frame, we expect that a RAP drives the system through the instantaneous ground state of $H_\text{eff}$ from a negative $t\rightarrow -\infty$ to a final positive detuning $t\rightarrow\infty$, passing through the minimum gap $\Omega_D$ at $t=0$. This is equivalent to the state transformation,
\beq
\ket{\Psi(-\infty)}=\ket{\Sstate}\rightarrow\ket{\Psi(\infty)}=e^{i(kx+\chi)}\ket{\Pstate}.
\eeq
The ion experiences a kick $k=\Delta/c$ in phase-space and performs the desired internal state transition up to a dynamical phase $\chi$ to be determined. 

When the excitation is incomplete, the action of the pulse on the laboratory frame that arrives at $t_0$ and ends at $t_f$ can be written as
\begin{equation}
    U(t_f,t_0;x,k,\xi) = e^{-\tfrac{i}{2}[\Delta t_f\pm kx]\sigma^z} e^{-i\chi\sigma^z}e^{-i\theta\sigma^x/2}e^{\tfrac{i}{2}[\Delta t_0\pm kx]\sigma^z}.
    \label{eq:single-pulse}
\end{equation}
Here, $\theta$ is the angle of rotation under the RAP pulse, which in the ideal limit is $\theta=\pi$. The passage also has an associated dynamical phase, $\chi$, and a phase $kx$ that accounts for the kick experienced by the ion.

\section{Non-overlapping pairs of pulses}

The dynamics under two pulses that do not overlap can be derived from the action of a single pulse\ \eqref{eq:single-pulse}, repeated twice, with a waiting time given by the time delay $T_d\to\pm\infty$ that is much longer than the FWHM of the chirped wavepacket. As before, we can abstract away the position dependence and write the two kicks as a composition of two partially successful rotations, with a relative phase dependent on the phase difference between pulses and the position of the ion
\begin{equation}
    U_\text{2kick} \propto e^{-i\chi \sigma^z/2}e^{-i\phi\sigma^z/2}e^{-i\chi\sigma^z/2},
\end{equation}
to be compared with
\begin{equation}
    U_\text{1kick}\propto\exp(-i\chi\sigma^x/2).
\end{equation}
We can compute the single-kick excitation probability from either protocol, obtaining
\begin{align}
    p_1 &= \left|\braket{\Pstate|U_\text{1kick}|\Sstate}\right|^2 = \sin\left(\frac\chi2\right)^2,\\
    p_2 &= \left|\braket{\Pstate|U_\text{2kick}|\Sstate}\right|^2 =  \cos\left(\frac\phi2\right)^2 \sin(\chi)^2.
\end{align}
We can slightly massage the latter formula using
\begin{equation}
\sin(\chi)^2 = 4 \sin\left(\frac\chi2\right)^2\left[1-\sin\left(\frac\chi2\right)^2\right],
\end{equation}
to find the relation $p_2 = 4\cos(\phi/2)^2 p_1(1-p_1).$

\section{RAP pulse interference}
\begin{figure*}
    \centering
    \includegraphics[width=18.2cm,height=5.5cm]{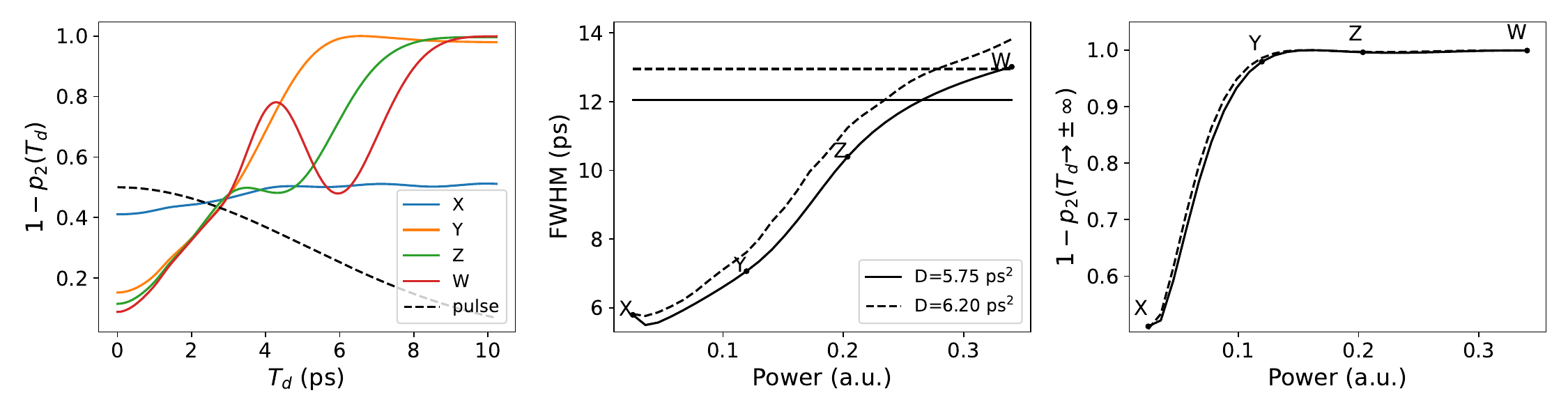}
        \caption{(left) Probability to recover the $\Sstate$ state after the interference of two pulses for four values of the pulse power, using $D=5.75$ $\mbox{ps}^2$. Profile Y corresponds to the fit of Fig.~\ref{AC} in the main text. (center) Excitation profile $p_2(T_d)$ full-width half-maximum (FWHM) at the broadest point, for $\sigma=1.153$ ps and two values of the chirp $D=5.75$ $\mbox{ps}^2$ (solid) and $6.20$ $\mbox{ps}^2$ (dashed). We show the FWHM of the chirped pulses in horizontal lines to illustrate how the interference profile is narrower than the original pulse and broadens with the pulse energy. (right) Probability of restoring the ion to the $\Sstate$ after two well-separate RAPs for similar values.}
    \label{fig:excitation-from-interference}
\end{figure*}
Let us consider two counter-propagating pulses that arrive at the same ion from opposite directions, with a time delay $T_d$. As before, we assume that both the duration of the pulses and the time delay are much shorter than the trapping period so that we can assume frozen dynamics for the ion. Similarly to the previous section, we model such a system with the Hamiltonian
\beq
\label{2pulses}
H=\frac{\Delta}{2}\sigma^z + \left[\Omega(t-x/c) + \Omega(t - T_d+x/c)\right] \sigma^+ + \mathrm{H.c.},
\eeq
where $\Omega(t) = \Omega_D f(t)$, given by Eq.~\eqref{chirp}, captures the chirped Gaussian profile of each pulse. The relative time delay $T_d$ is controlled by an adjustable path length in one of the interferometric arms, and it may be affected by the ion's position through the $\pm x/c$ term.

The interference between pulses gives rise to a non-trivial excitation pattern which requires the full simulation of the Schrödinger equation with the Hamiltonian~\eqref{2pulses}. After this simulation, we will find the ion back in the $\Sstate$ with a probability $p_2(T_d)$. This probability already had a nonlinear dependency on the power of the chirped RAP for very separated pulses, as evidenced in Fig.\ \ref{fig:rap}. However, now that the pulses can interfere, this nonlinear dependency is further enhanced by the appearance of a chirped optical lattice that shakes the ion, invalidating the quasi-adiabatic picture of the rapid adiabatic passage.

This dynamics is illustrated in Fig.\ \ref{fig:excitation-from-interference}(left). This plot shows the probability of recovering the $\Sstate$ after two counter-propagating pulses kick the ion separated a time $T_d$. These excitation profiles have been averaged over all possible relative pulse phases, for otherwise, the outcome would be a dense plot with very narrow oscillations, enveloped by the curves in Fig.\ \ref{fig:excitation-from-interference}. As for the outcome of the numerical experiment, notice how when the power of the pulses is low, or the two pulses strongly overlap and interfere destructively, the two pulses cannot restore the ion to the $\Sstate$ state. However, the shape of these curves is not exactly one of the incoming pulses. Instead, they are Gaussian-like profiles that broaden with the pulses' energy and become distorted for stronger powers.

Despite this apparent complexity, our numerical simulations suggest a proportionality between the incoming pulse width and the FWHM width of the two-pulse interference profile. However, this relation is power-dependent, as evidenced in Fig.\ \ref{fig:excitation-from-interference}(center), which means that we need to fit the original pulse width $\sigma$, the GDD chirping factor $D$ and the pulse power $I\propto|\Omega_D|^2$ simultaneously. To improve the accuracy of this reconstruction, various strategies can be used: (i) gather an independent, high-accuracy calibration of the chirping $D$; (ii) gather as many points as possible from the $p_2(T_d)$ curve and (iii) include in the fit the curves of single RAP pulses $p_1$ or the curve $p_2(T_d\to+\infty)$ for well-separated pulses, as a function of power (e.g., Fig.\ \ref{fig:rap}
in the manuscript). Combining these techniques will significantly reduce the uncertainty of the fit.

Unfortunately, in this experiment, we could not combine the fits from Fig.\ \ref{fig:rap} and Fig.\ \ref{AC}, because of fluctuations in the frequency doubling processes. The lack of this combined fitting strategy results in large uncertainty in the estimation of the parameters $D$, $I$, and $\sigma$, which is well below the potential limits of this powerful technique.

\section{Spin-echo experiment}
So far, this discussion refers to the interference of two pulses on a single ion. What about the spin-echo experiment? Remember that this interferometry experiment was calibrated for a time delay of exactly one trap period. In this situation, we can prove that the contrast of the interference fringes is proportional to the probability $p_2(T_d)$ analyzed in the previous section.

The initial state of the ion in the Ramsey interference experiment is
\begin{equation}
    \ket{\psi_0} = \frac{1}{\sqrt{2}}(\ket{\Sstate} + \ket{\Dstate})\otimes\ket{\alpha_0}.
\end{equation}
We have now introduced the $\Dstate$ state that encodes the qubit's excited state. The first two pulses transform the $\Sstate$ state into a coherent superposition of $\Sstate$ and $\Pstate$ states
\begin{multline*}
    \ket{\psi_1} = \frac{1}{\sqrt{2}}\left(p_2^{1/2}e^{i\phi}\ket{\Sstate}\ket{\alpha_1} + (1-p_2)^{1/2}\ket{\Pstate}\ket{\alpha_2}\right)\\ +\frac{1}{\sqrt{2}}\ket{\Dstate}\otimes\ket{\alpha_0},
\end{multline*}
with different motional states and the excitation probability $p_2$ we discussed in the pulse interference section above. Naturally, the $\Pstate$ state will decay rapidly compared to the motional speed. We can approximate the resulting mixed state as an incoherent superposition
\begin{equation}
    \rho_3 = \ket{\psi_3}\bra{\psi_3} + (1 - p_2)\ketbra{\Sstate}{\Sstate}\otimes\rho_\text{kick},
\end{equation}
of a pure state
\begin{equation}
    \ket{\psi_3} = \frac{1}{\sqrt{2}}\left(P_2^{1/2}e^{i\phi}\ket{\Sstate}\ket{\alpha_1} + \ket{\Dstate}\ket{\alpha_0}\right),
\end{equation}
and a mixed motional state $\rho_\text{kick}$ describing our ignorance of the motional state of the ion after spontaneous emission.

Waiting for half a trap period, swapping $\Sstate$ and $\Dstate$ states, and waiting for another half-a-trap period transforms this density matrix into
\begin{equation}
    \rho_4 = \ket{\psi_4}\bra{\psi_4} + (1 - P_2)\ketbra{\Dstate}{\Dstate}\otimes\rho_\text{kick},
\end{equation}
with the pure state
\begin{equation}
    \ket{\psi_4} = \frac{1}{\sqrt{2}}\left(P_2^{1/2}e^{i\phi}\ket{\Dstate}\ket{\alpha_1} + \ket{\Sstate}\ket{\alpha_0}\right),
\end{equation}

As before, the second pair of pulses will operate on the $\Sstate$ state, transforming it into a superposition of $\Sstate$ and $\Pstate$, with some changes in the motional state that we assume are identical. The resulting density matrix is
\begin{multline*}
    \rho_5 = P_2\ket{\psi_5}\bra{\psi_5} + \frac{1-P_2}{2}\\
    \left(\ketbra{\Sstate}{\Sstate}+\ketbra{\Dstate}{\Dstate}\right)\otimes\rho_\text{kick},
\end{multline*}
with the pure state
\begin{equation}
    \ket{\psi_5} = \frac{1}{\sqrt{2}}\left(e^{i\xi}\ket{\Dstate} + \ket{\Sstate}\right)\ket{\alpha_1}.
\end{equation}
This state has some relative phase $\xi = \theta + \delta\xi$. The spin-echo contrast measurement tries different values of the phase $\xi$ and studies the difference between the maximum and the minimum values. In this scenario, the probability $(1-p_2)/2$ value is the background, and the visibility of the fringes is given by $p_2 \times \braket{\cos(\delta\xi)^2}$. This value includes the fluctuations in the relative phases of pairs of pulses $\delta\xi$, but it may also be degraded by the lack of precise repeatability in the changes of motional state---e.g., if the moment imparted onto the ion changes slightly due to timing errors.

According to this theory, the data in Fig.\ \ref{AC}a can be fitted to the numerically predicted interference profiles $p_2(T_d, \Omega_D, \sigma, D)$ for an experiment with only two counter-propagating pulses, separated a time $T_d$, with Rabi frequency $\Omega_D$, pulse width $\sigma$ and chirping $D$. In doing so, we find the solid curve from Fig.\ \ref{AC}, which gives an estimate of the pulse width $t_\text{FWHM}=12$ ps.

\bibliography{mybib.bib}

\end{document}